\documentclass[preprint,floatfix] {revtex4} 
 
\usepackage{graphicx}
\usepackage{subfigure}
\usepackage{amsmath}
\begin{document}

\title{Confinement in 3D polynomial oscillators through a generalized pseudospectral method}
\author{Amlan K. Roy}
\altaffiliation{Email: akroy@iiserkol.ac.in, akroy6k@gmail.com, Ph: +91-3473-279137, Fax: +91-33-25873020.}
\affiliation{Division of Chemical Sciences,   
Indian Institute of Science Education and Research Kolkata, 
Mohanpur Campus, Nadia, 741252, India}

\begin{abstract}
Spherical confinement in 3D harmonic, quartic and other higher oscillators of even order is studied. The generalized
pseudospectral method is employed for accurate solution of relevant Schr\"odinger equation in an \emph{optimum,
non-uniform} radial grid. Eigenvalues, 
eigenfunctions, position expectation values, radial densities in \emph{low and high-lying} states are presented in case of 
\emph{small, intermediate and large} confinement radius. The \emph{degeneracy breaking} in confined situation as well as 
correlation in its \emph{energy ordering} with respect to the respective unconfined counterpart is discussed. For all 
instances, current results agree 
excellently with best available literature results. Many new states are reported here for first time. In essence, a
simple, efficient method is provided for accurate solution of 3D polynomial potentials enclosed within spherical
impenetrable walls. 


\end{abstract}
\maketitle

\section{Introduction}
Ever since the study of hydrogen atom inside an impenetrable spherical enclosure \cite{michels37}, 
confinement of quantum systems has witnessed many burgeoning activity in past few decades. This 
work was motivated by an attempt to understand effects of extreme pressure on electronic states. 
In such extremely small spatial dimensions, many fascinating phenomena occur in a quantum confined system, 
relative to the respective free system. Some of these relate to, e.g., thermodynamic properties in 
non-ideal gases, anharmonic effects in solids, impurity binding in quantum wells, partially ionized 
plasmas, mesoscopic scale artificial structures, etc. Gradually their importance was realized in 
various potential applications in physics and chemistry, such as quantum wells, quantum wires, quantum 
dots, etc. They have also relevance in the development of numerous nano-sized circuits including quantum computers. 
Usually, their unique physical, chemical properties are attributed to complex energy changes associated with them. 
Most widely studied quantum systems are particle in a box, harmonic oscillator, H atom, He and other 
many-electron atoms, as well as H$_2^+$, H$_2$ and other molecules. The literature is vast; here we mention
a few representatives \cite{fowler84, froman87, yngve88, jaskolski96, connerade00, buchachenko01, 
gravesen05, heiss05}.

Confinement problem in 1D confined harmonic oscillator (CHO) has been theoretically investigated by a number of workers over a 
long period of time \cite{vawter68,navarro80,fernandez81,arteca83,taseli93,vargas96,sinha99,aquino01,campoy02}. 
For example, the effect of finite boundaries on a 1D CHO inside a potential enclosure was studied by means of WKB method 
\cite{vawter68}. Perturbative, asymptotic and Pad\'e approximant solutions for boxed-in harmonic and inverted
oscillators were reported \cite{navarro80}. Eigenfunctions, eigenvalues in such 1D potentials were also 
obtained by means of a hypervirial method \cite{fernandez81,arteca83}. Accurate energies of 1D polynomial potentials 
having single and multiple wells inside a box were calculated through a Rayleigh-Ritz variation method by using
a basis set of trigonometric functions \cite{taseli93}. A robust, strongly convergent numerical method has
been proposed for harmonic, quartic and sextic oscillators \cite{vargas96}. Also, the trapped harmonic and quartic
oscillators have been studied by a WKB method with appropriate boundary condition \cite{sinha99}. This employed a modified 
airy function method that ultimately leads to a modified Bohr-Sommerfeld-type quantization rule. Eigenvalues, dipole moments, 
Einstein coefficients of CHO have been considered by perturbation method \cite{aquino01} as well. Symmetric and asymmetric confinement 
was studied in \cite{campoy02} by a power series expansion and numerical method. 

In parallel to the 1D problem mentioned above, some attention was been paid for similar studies in 2D, 3D and higher 
dimension, however with lesser intensity \cite{fernandez81a,navarro83,marin91,aquino97,taseli97a,taseli97,sinha03,filho03,sen06,
montgomery07,jaber08,stevanovic08a,stevanovic08, montgomery10,serrano13}. Several unique phenomena occur in such systems, 
especially related to \emph{simultaneous, incidental and interdimensional} degeneracy \cite{sen06,jaber08,stevanovic08a,
stevanovic08,montgomery07}. In one of the earliest studies a hypervirial treatment was suggested for multidimensional isotropic 
bounded oscillators, including 3D CHO \cite{fernandez81a}. The 2D and 3D CHOs enclosed in circular and spherical boxes were 
considered in \cite{navarro83} by a Rayleigh-Schr\"dinger perturbative expansion as well as Pad\'e approximant solution. 
A variational method \cite{marin91} was proposed where the trial function was constructed as a product of ``free" solutions of 
corresponding Schr\"odinger equation. However, the first systematic and accurate ground, excited states of 3D CHO (in terms 
of energies, eigenfunctions, spatial expectation values) with respect to box size were obtained in variational calculation 
of \cite{aquino97}. Later, results were also obtained by a WKB method \cite{sinha03}. A super-symmetric method 
\cite{filho03} also offered reasonably accurate results. The $N$-dimensional CHO enclosed in an impenetrable spherical cavity 
was also computed by searching for the roots of hypergeometric function $_1F_1 (a,b,z)$ \cite{jaber08}. Later, high-precision 
energies for $N$-dimensional CHO were presented by exploiting certain properties of hypergeometric function within the MAPLE 
computer algebra system \cite{montgomery10}. Very recently, a combination of WKB method and a proper quantization rule has been
advocated for CHO energy levels \cite{serrano13}. Energy characteristics of a 2D isotropic CHO, enclosed in a 
symmetric box were followed through annihilation and creation operators coupled with the infinitesimal operators of 
SU(2) group \cite{stevanovic08}. Likewise, characteristic features in energy spectrum of 3D CHO were established by invoking 
analytical properties of Bessel and confluent hypergeometric functions. However, the most extensively studied confined system 
is H atom. A vast amount of reference exists on the subject emphasizing different kinds of confinement and theoretical 
methods \cite{aquino95,burrows06,aquino07,ciftci09}.

In present communication, we are interested in the confinement studies in 3D quantum polynomial oscillators. It may be 
noted that while all these above methods produce reasonably good-quality results, only two \cite{aquino97, montgomery10} 
of these offer energies with ten or higher significant figure accuracy for 3D oscillators. Moreover, there is a lack of 
good results in the \emph{low radius of confinement}, and \emph{higher excited states}. Further, while a substantial
amount of accurate results exist for enclosed H atom problem, same for 3D CHO is much less and almost none for other polynomial
potentials. We will use the generalized pseudospectral (GPS) method for this purpose, which in recent years, has been 
successfully applied to a wide range of \emph{unconfined} physical and chemical systems. Some 
of these are spiked harmonic oscillator, Hulth\'en, Yukawa, logarithmic, rational, power-law potentials, ro-vibrational 
levels in molecules as well as ground and Rydberg states in atoms, etc. \cite{roy04, roy04a, roy04b, roy05, roy05a, sen06, 
roy08}. Only in \cite{sen06}, however, this was employed for confined quantum states. There, a few low-lying states such 
as $1s, 2p, 1d$ of 3D CHO were examined; additionally some degeneracy conditions were established. 
However a detailed performance of the GPS method in case of \emph{confined} systems has not been made so far. In this 
work, we want to test its validity and relevance in such a context, which could extend its domain of applicability to 
larger set of problems. To this end, eigenvalues, eigenfunctions, expectation values are reported for low, high states. A detailed
variation of eigenvalues with box size is also presented. Correlation between energies in \emph{free} and respective \emph{confined} 
system is discussed. Further, a similar study is made on quartic oscillator, which was not done before. Finally we consider
confinement in even-degree polynomial oscillators (up to 20) in terms of changes in ground-state energy with radius of 
confinement. Section II gives a brief outline of the method used. Results are discussed in section III. We conclude with a 
few remarks in section IV. 

\section{Method of calculation}             
The GPS method has been found to be very successful for various physical and chemical systems as evidenced by the following
applications \cite{roy04, roy04a, roy04b, roy05, roy05a, sen06, roy08}. Therefore here we highlight only the essential aspects. 
Unless otherwise mentioned, atomic unit is employed. 

The single-particle, non-relativistic time-independent radial Schr\"odinger equation can be written in following form (the 
problem is separable in radial and angular variables), 
\begin{equation}
\left[-\frac{1}{2} \ \frac{\mathrm{d^2}}{\mathrm{d}r^2} + \frac{\ell (\ell+1)} {2r^2}
+v(r) +v_c(r) \right] \psi_{n,\ell}(r) = E_{n,\ell}\ \psi_{n,\ell}(r)
\end{equation}
where $v(r)$ is a polynomial potential of degree $2K$ containing only the dominant term, i.e., $v(r) = \frac{1}{2} r^{2K}$. 
In this work we consider $K=1,2,3, \cdots , 10$, while $n$, $\ell$ signify usual radial and angular momentum quantum 
numbers respectively. Last term in square bracket represents the confinement potential ($r_c$ is the radius of confining 
spherical box),
\begin{equation} v_c(r) = \begin{cases}
+\infty \ \ \ \ r> r_c  \\
0,  \ \ \ \ \ \ \ r \leq r_c   \\
 \end{cases} 
\end{equation}
This equation needs to be solved with Dirichlet boundary condition $\psi_{n, \ell}(0)=\psi_{n,\ell}(r_c)=0$.

A very distinctive feature of this approach is that it allows one to work in a \emph{nonuniform}, \emph{optimal} spatial 
discretization; a coarser mesh at larger $r$ and a denser mesh at smaller $r$, while maintaining similar accuracies at both 
regions. Thus, compared to many other methods, much smaller number of grid point often suffices to recover 
very good accuracy for low and high states. 

At first, a function $f(x)$ defined in an interval $x \in [-1,1]$ is approximated by an N-th order polynomial $f_N(x)$,
\begin{equation}
f(x) \cong f_N(x) = \sum_{j=0}^{N} f(x_j)\ g_j(x),
\end{equation}
so that the approximation is \emph {exact} at \emph {collocation points} $x_j$, i.e., $ f_N(x_j) = f(x_j).$
In the Legendre pseudospectral method used here, $x_0=-1$, $x_N=1$, while  $x_j (j=1,\ldots,N-1)$ are obtained from roots of 
first derivative of Legendre polynomial $P_N(x)$ with respect to $x$, i.e., $P'_N(x_j) = 0.$
Cardinal functions, $g_j(x)$ in Eq.~(3) satisfy the relation, $g_j(x_{j'}) = \delta_{j'j}$. Next, one can map the 
semi-infinite domain $r \in [0, \infty]$ onto finite domain $x \in [-1,1]$ by a transformation $r=r(x)$ and introduce a
nonlinear algebraic mapping of following form, 
\begin{equation}
r=r(x)=L\ \ \frac{1+x}{1-x+\alpha},
\end{equation}
where L and $\alpha=2L/r_{max}$ are two adjustable mapping parameters. Now, introducing a transformation of the type
$\psi(r(x))=\sqrt{r'(x)} f(x), $
followed by a symmetrization procedure leads to a \emph {symmetric} matrix eigenvalue problem which can be easily solved by 
standard available routines (NAG libraries used) to yield accurate eigenvalues and eigenfunctions. 

Sample calculations were performed for a sufficiently large sets of mapping parameters to monitor the accuracy and reliability 
of present method, so as to produce ``stable'' results with respect to reference results. This lead to a choice of 
$\alpha=25, N=200$, which appeared to be satisfactory for the problem at hand. Results in various tables are reported only up 
to the precision that maintained stability and they are {\em truncated} rather than {\em rounded-off}. Thus, they may be 
considered as correct up to the decimal place given.    

\begingroup
\squeezetable
\begin{table}
\caption {\label{tab:table1}Energies (a.u.) of 3D CHO for $n=\{0,1\}; \ell=\{0,1,2\}$ states. PR means Present Result.} 
\begin{ruledtabular}
\begin{tabular}{l|ll|ll}
$r_c$  & E$_{0,0}$ (PR)   & E$_{0,0}$ (Literature) &  E$_{0,1}$ (PR)  &  E$_{0,1}$ (Literature)   \\
\hline
0.1    & 493.48163346     &  493.481632\footnotemark[1]$^,$\footnotemark[2] 
       & 1009.5383008     &  1009.538302\footnotemark[1]$^,$\footnotemark[2]         \\
0.3    & 54.843855432     &  54.843855\footnotemark[1]$^,$\footnotemark[2] 
       & 112.18757119     &  112.187571\footnotemark[1]$^,$\footnotemark[2]         \\
0.5    & 19.774534180     & 19.774534\footnotemark[1]$^,$\footnotemark[2]$^,$\footnotemark[3],19.774534179\footnotemark[4]$^,$\footnotemark[5]
       & 40.428276496     &  40.428277\footnotemark[1]$^,$\footnotemark[2],40.428276497\footnotemark[4],40.428276496\footnotemark[5]   \\
1.0    & 5.0755820154     & 5.0755820\footnotemark[3],5.0755820152\footnotemark[4]$^,$\footnotemark[5]$^,$\footnotemark[6],4.98501\footnotemark[7] 
       & 10.282256939     & 10.282256939\footnotemark[4]$^,$\footnotemark[5]$^,$\footnotemark[6],10.13941\footnotemark[7] \\ 
1.5    & 2.5049761786     & 2.502471\footnotemark[1],2.504976\footnotemark[2],2.5049761\footnotemark[3],
       & 4.9035904193     & 4.901782\footnotemark[1],4.903590\footnotemark[2],            \\
       &                  & 2.5049761785\footnotemark[4]$^,$\footnotemark[6],2.54357\footnotemark[7]  
       &                  & 4.9035904194\footnotemark[4]$^,$\footnotemark[6],4.87758\footnotemark[7]                                                 \\
2.0    & 1.7648164388     & 1.732515\footnotemark[1],1.764816\footnotemark[2],  
       & 3.2469470987     & 3.224167\footnotemark[1],3.246947\footnotemark[2],   \\
       &                  & 1.7648087\footnotemark[3],1.7648164387\footnotemark[4]$^,$\footnotemark[5]   
       &                  & 3.2469470987\footnotemark[4]$^,$\footnotemark[5]$^,$\footnotemark[6],3.31861\footnotemark[7]                     \\
3.0    & 1.5060815273     & 1.544195\footnotemark[1],1.506082\footnotemark[2],1.5060815272\footnotemark[4]
       & 2.5312924665     & 2.688286\footnotemark[1],2.531292\footnotemark[2],2.5312924666\footnotemark[4]  \\
4.0    & 1.5000146030     & 1.504181\footnotemark[1],1.5000015\footnotemark[2],1.50000146030\footnotemark[4]
       & 2.5001437781     & 2.517144\footnotemark[1],2.500144\footnotemark[2],2.5001437781\footnotemark[4]   \\
5.0    & 1.5000000037     & 1.500581\footnotemark[1],1.500000\footnotemark[2],1.5000000036\footnotemark[4]$^,$\footnotemark[5]
       & 2.5000000584     & 2.502500\footnotemark[1],2.500000\footnotemark[2],2.5000000584\footnotemark[4]$^,$\footnotemark[5] \\
6.0    & 1.5000000000     & 1.5000000000\footnotemark[4]                 
       & 2.5000000000     & 2.5000000000\footnotemark[4]      \\
\hline
       & E$_{0,2}$ (PR)    & E$_{0,2}$ (Literature) &  E$_{1,0}$ (PR)  &  E$_{1,0}$ (Literature)   \\
\hline
0.1    & 1660.8752892     &                       
       & 1973.9224834     &                             \\
0.3    & 184.56119633     &                       
       & 219.33897241     &                             \\
0.5    & 66.489756534     & 66.489756534\footnotemark[4]     
       & 78.996921147     & 78.996921150\footnotemark[4]$^,$\footnotemark[5]                                          \\
1.0    & 16.827777109     & 16.827777109\footnotemark[4] 
       & 19.899696502     & 19.899696501\footnotemark[4]$^,$\footnotemark[5]$^,$\footnotemark[6],19.78300\footnotemark[7]  \\
1.5    & 7.8717304875     & 7.8717304877\footnotemark[4] 
       & 9.1354220880     & 9.1354220876\footnotemark[4]$^,$\footnotemark[6],9.10021\footnotemark[7]                 \\
2.0    & 5.0100408655     & 5.0100408656\footnotemark[4] 
       & 5.5846390792     & 5.5846390790\footnotemark[4]$^,$\footnotemark[5]$^,$\footnotemark[6],5.59685\footnotemark[7]   \\
3.0    & 3.5982476989     & 3.5982476989\footnotemark[4]
       & 3.6642196451     & 3.6642196450\footnotemark[4]$^,$\footnotemark[6],4.01642\footnotemark[7]                                \\
4.0    & 3.5008420737     & 3.5008420738\footnotemark[4]
       & 3.5016915386     & 3.5016915385\footnotemark[4]                                                        \\
5.0    & 3.5000005566     & 3.5000005567\footnotemark[4]
       & 3.5000012215     & 3.5000012214\footnotemark[4]$^,$\footnotemark[5]                                   \\
6.0    & 3.5000000000     & 3.5000000003\footnotemark[4]   
       & 3.5000000001     & 3.5000000000\footnotemark[4]                                                       \\
\hline
       & E$_{1,1}$ (PR)   & E$_{1,1}$ (Literature) &  E$_{1,2}$ (PR)  &  E$_{1,2}$ (Literature)   \\
\hline
0.1    & 2983.9775336     &                          
       & 4135.9634334     &                                                                \\
0.3    & 331.56849477     &  
       & 459.56818797     &                                                                \\
0.5    & 119.40244526     & 119.40244525\footnotemark[4],119.40244526\footnotemark[5]
       & 165.48541838     &                                                                \\
1.0    & 30.013487593     & 30.013487592\footnotemark[4],30.013487591\footnotemark[5]$^,$\footnotemark[6],29.88263\footnotemark[7]
       & 41.547472167     & 41.547472166\footnotemark[6],41.40000\footnotemark[7]                                       \\
1.5    & 13.653740893     & 13.653740893\footnotemark[4],13.653740892\footnotemark[6],13.60644\footnotemark[7]   
       & 18.805042927     & 18.805042927\footnotemark[6],18.74733\footnotemark[7]                                       \\
2.0    & 8.1595288818     & 8.1595288816\footnotemark[4]$^,$\footnotemark[5]$^,$\footnotemark[6],8.15237\footnotemark[7] 
       & 11.093419194     & 11.093419194\footnotemark[6],11.07445\footnotemark[7]                                       \\ 
3.0    & 4.9138976907     & 4.9138976907\footnotemark[4]$^,$\footnotemark[6],5.05725\footnotemark[7]
       & 6.3076308117     & 6.3076308118\footnotemark[6],6.37769\footnotemark[7]                                        \\
4.0    & 4.5083304309     & 4.5083304308\footnotemark[4]    
       & 5.5286815696     &                                                                            \\
5.0    & 4.5000105731     &  4.5000105730\footnotemark[4]$^,$\footnotemark[5]
       & 5.5000647985     &                                                                            \\
6.0    & 4.5000000011     &  4.5000000008\footnotemark[4]                  
       & 5.5000000098     &                                                                            \\  
\end{tabular}
\end{ruledtabular}
\begin{tabbing}
$^{\mathrm{a}}$Ref.~\cite{navarro83}. \hspace{15pt}  \=
$^{\mathrm{b}}$As quoted in \cite{navarro83}. \hspace{15pt}  \=
$^{\mathrm{c}}$Ref.~\cite{fernandez81a}. \hspace{15pt}  \= 
$^{\mathrm{d}}$Ref.~\cite{aquino97}. \hspace{15pt} \=
$^{\mathrm{e}}$Ref.~\cite{montgomery10}. \hspace{15pt} \= 
$^{\mathrm{f}}$As quoted in \cite{serrano13}. \hspace{15pt} \= 
$^{\mathrm{g}}$Ref.~\cite{serrano13}. 
\end{tabbing}
\end{table}
\endgroup

\begingroup
\squeezetable
\begin{table}
\caption {\label{tab:table2}Eigenvalues (in a.u.) of $n=8, \ell=\{0-8\}$ states of 3D CHO as function of $r_c$.} 
\begin{ruledtabular}
\begin{tabular}{lllllll}
$\ell$  & $r_c=0.5$ & $r_c=1.5$       &  $r_c=2.5$     &  $r_c=3.5$      &   $r_c=4.5$     &  $r_c=10$   \\
\hline
0   &  1598.9175016  &  178.02733028   &  64.998075416   &  34.693293571   &  23.225703703   &  17.500000000  \\
1   &  1777.5046290  &  197.87206375   &  72.146061470   &  38.344628587   &  25.432849261   &  18.499999999  \\
2   &  1961.9446659  &  218.36830761   &  79.531522574   &  42.122176083   &  27.724249606   &  19.499999999  \\
3   &  2152.2122548  &  239.51291591   &  87.152525538   &  46.023630724   &  30.096639843   &  20.500000000  \\
4   &  2348.2769485  &  261.30226934   &  95.007188871   &  50.047065889   &  32.547442956   &  21.500000000  \\
5   &  2550.1056822  &  283.73252093   &  103.09370240   &  54.190838285   &  35.074586281   &  22.500000000  \\
6   &  2757.6642785  &  306.79974464   &  111.41033572   &  58.453522134   &  37.676374299   &  23.499999999  \\
7   &  2970.9183755  &  330.50002581   &  119.95544047   &  62.833862466   &  40.351398586   &  24.499999999  \\  
8   &  3189.8339986  &  354.82951614   &  128.72744924   &  67.330741133   &  43.098472783   &  25.499999999  \\

\end{tabular}
\end{ruledtabular}
\end{table}
\endgroup

\section{Results and Discussion}
At first, in Table I, we present first six eigenvalues corresponding to radial quantum numbers $n=0$ and 1 having 
$\ell=0,1,2,$ of 3D isotropic CHO in a spherical box with impenetrable walls. Depending on the box size, confinement 
can be classified in three distinct regions, namely, \emph{small, intermediate} and \emph{large} $r_c$. Ten cage radii 
were chosen for this purpose.  First definitive result of the lowest two states were presented through a
Rayleigh-Schr\"odinger-type perturbation expansion having free-particle in a box as the unperturbed system \cite{navarro83}. 
They also computed these states employing various Pad\'e approximations. Here, we quote their $P[1/5]$ energies
in third column, while perturbation energies are omitted to save space. Perturbation and Pad\'e eigenvalues differ 
amongst each other significantly, especially when the system is enclosed in a larger box (larger $r_c$). Their numerical
eigenvalues obtained by diagonalizing the Hamiltonian matrix in a basis of free particle-in-a-box eigenfunctions 
appear to be superior to above both, and referred here. While these are decent initial estimates, present GPS
energies are significantly improved over these. Confinement in the intermediate region was studied for ground state by 
means of a hypervirial method \cite{fernandez81a}. In this case, the accuracy generally becomes less precise as $r_c$ 
increases; for low $r_c$, these are more or less of similar quality as in \cite{navarro83}. Quite accurate eigenvalues  
for all states except E$_{1,2}$ (in the range $r_c > 0.5$) have been reported through variational procedure \cite{aquino97}. 
Here the two integers in subscript denote $n$, $\ell$ quantum numbers 
respectively. Current energies are in excellent agreement with these values for whole range of $r_c$; in several occasions
exactly reproducing those of the reference. However the most accurate energies are those published in \cite{montgomery10}; 
these are obtained as roots of a transcendental equation written in terms of confluent hypergeometric functions. 
The authors used familiar MAPLE algebra program to achieve impressive (as high as up to 100-digit) accuracy in eigenvalues. 
These are available for $n=\ell=0,1$ states at selected $r_c$. Here also, the GPS results offer excellent agreement with 
these energies. Very recently some of these states are also calculated by a combination of quantization rule and WKB
method \cite{serrano13}. These are also produced here for comparison. Obviously in all instances, isotropic free harmonic 
oscillator (FHO) energies are recovered as $r_c \rightarrow \infty$.  

\begin{figure}
\begin{minipage}[c]{0.40\textwidth}
\centering
\includegraphics[scale=0.45]{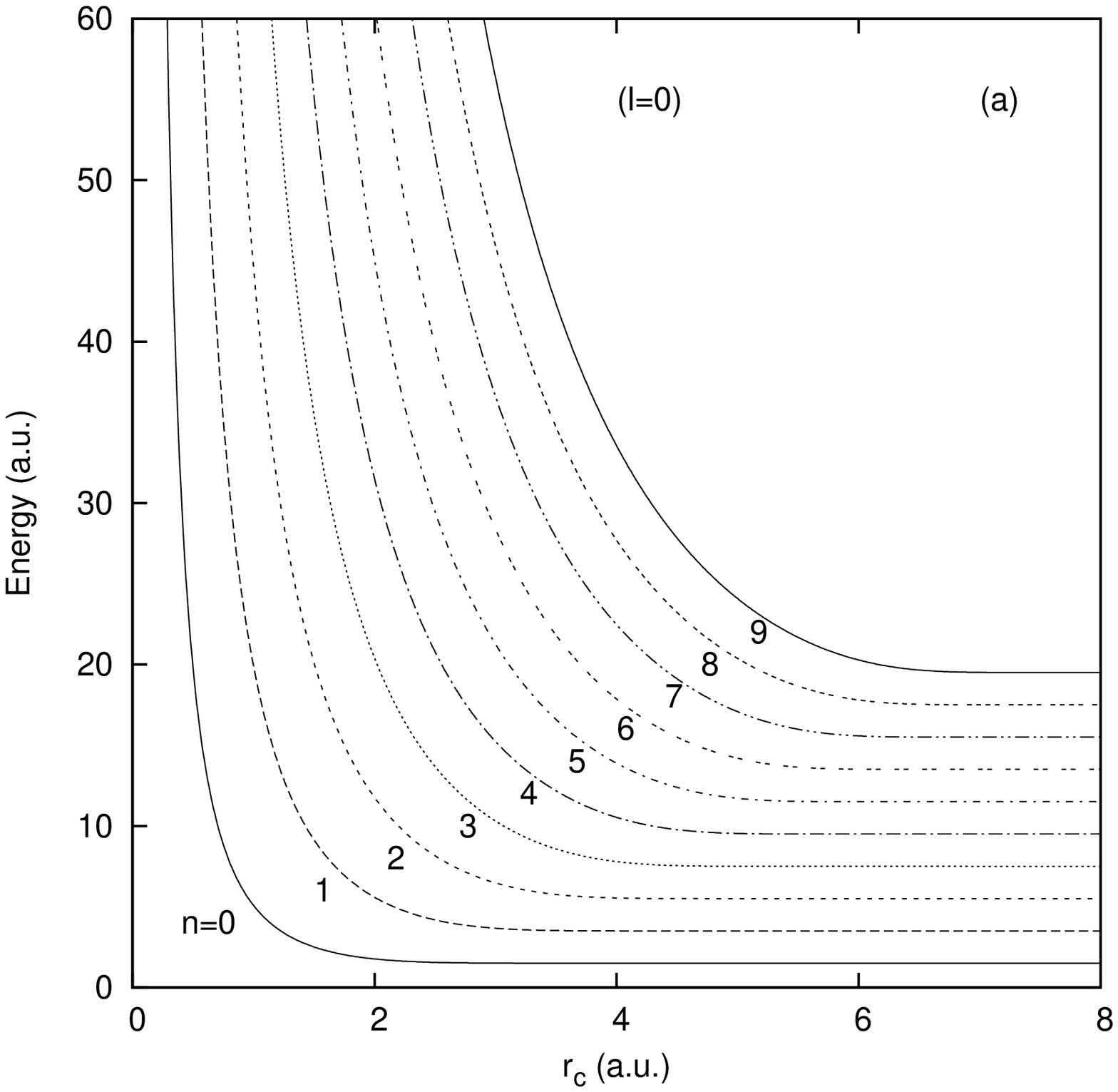}
\end{minipage}%
\hspace{0.5in}
\begin{minipage}[c]{0.40\textwidth}
\centering
\includegraphics[scale=0.45]{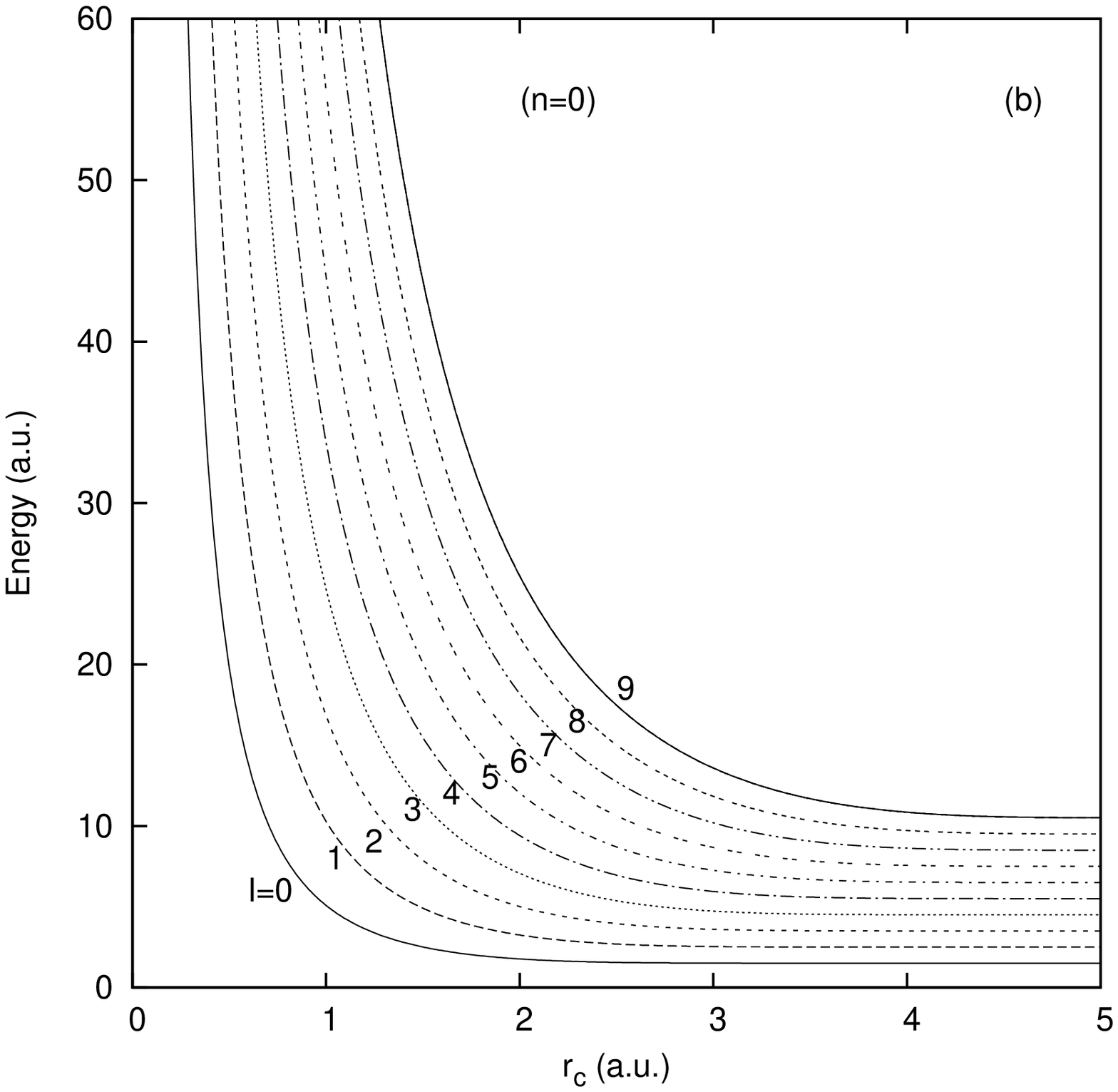}
\end{minipage}%
\caption{Energy changes (a.u.) in 3D isotropic CHO for (a) $l=0; n=0-9$ and (b) $n=0; l=0-9$ levels 
respectively, as function of confining radius.}
\end{figure}

Next we proceed in Table II for sample results on high-lying states. In order to demonstrate the viability and efficiency
in such situations, nine states (having $\ell=0-8$) corresponding to $n=8$ are given. As in previous table, here
again we cover the entire domain of confinement choosing six representative $r_c$ values, \emph{viz.}, 0.5, 1.5, 2.5, 3.5, 
4.5 and 10 a.u. respectively. As evident from above discussion, while a decent number of high-quality references exist
for confinement in low-lying states, there is a lack of such studies for higher excitations. Thus no literature values 
could be quoted for any of these. We hope that they would be helpful in future calibration of other methodologies. Once 
again, as box radius increases, energies of bounded system approach towards the 3D isotropic FHO eigenvalues.  

\begin{figure}
\begin{minipage}[c]{0.28\textwidth}
\centering
\includegraphics[scale=0.32]{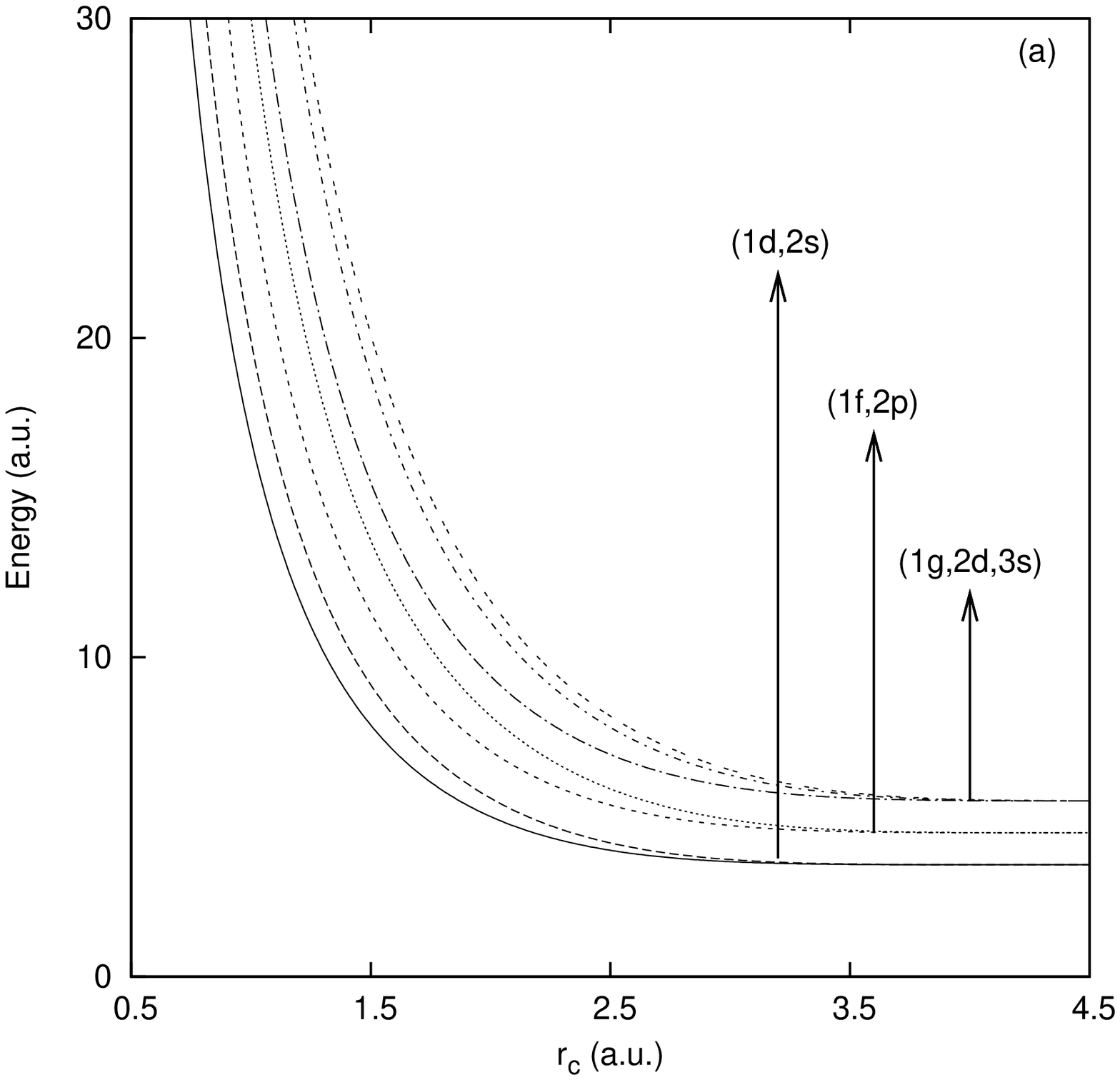}
\end{minipage}
\hspace{0.3in}
\begin{minipage}[c]{0.28\textwidth}
\centering
\includegraphics[scale=0.32]{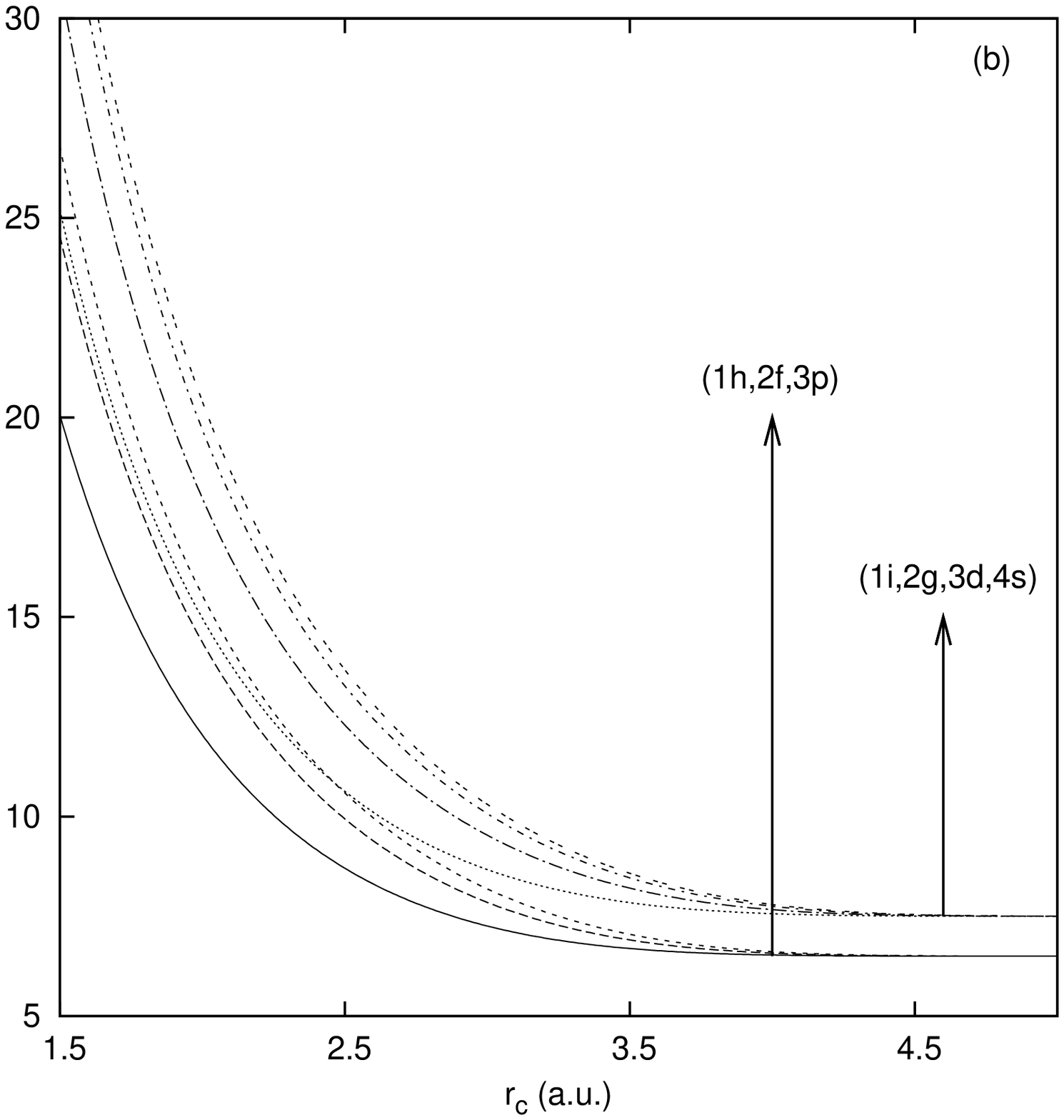}
\end{minipage}
\hspace{0.3in}
\begin{minipage}[c]{0.28\textwidth}
\centering
\includegraphics[scale=0.32]{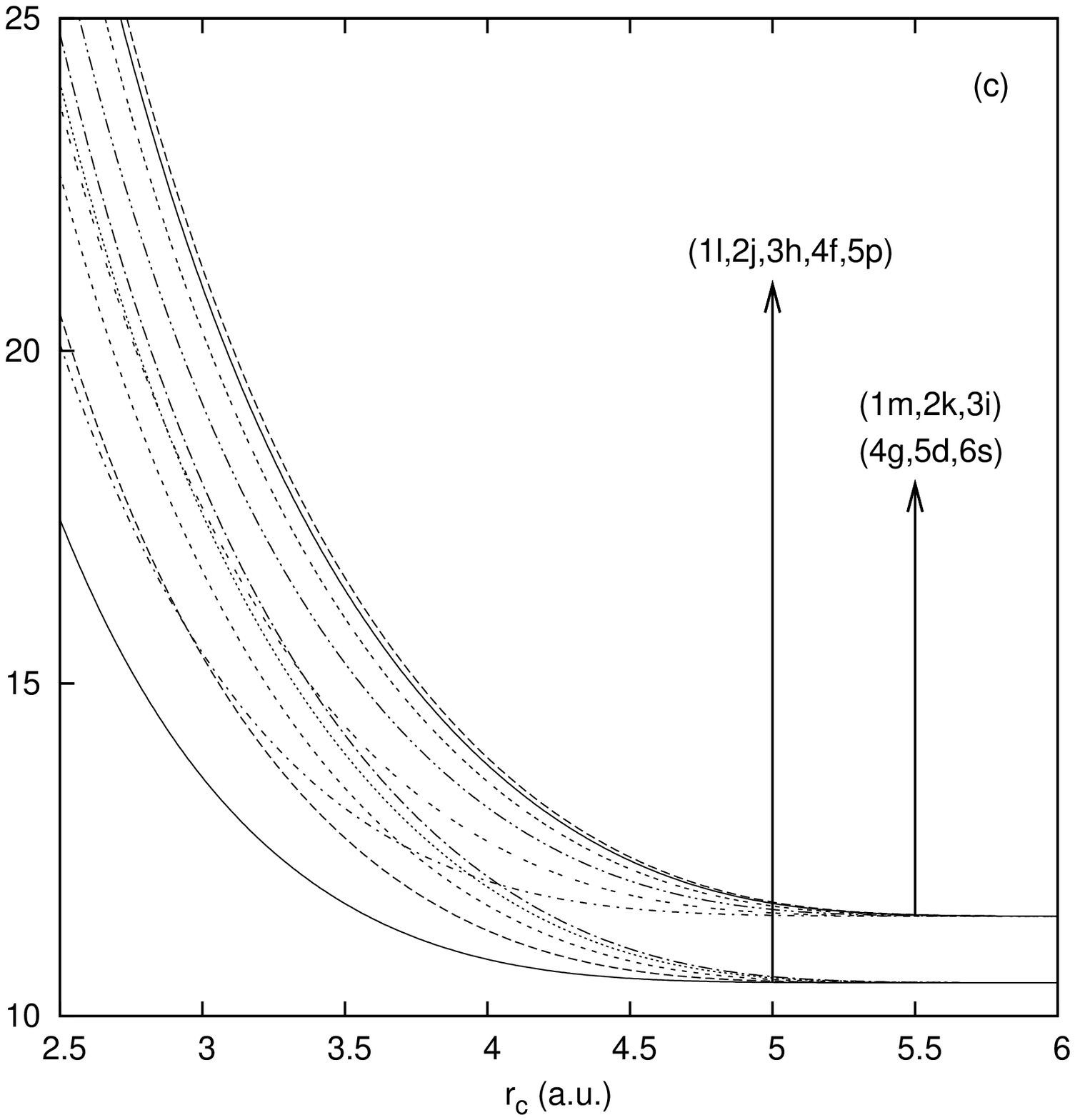}
\end{minipage}
\caption{Energy eigenvalues (a.u.) of (a) $(1d, 2s)$, $(1f, 2p)$, $(1g,2d,3s)$ (left), (b) $(1h,2f,3p)$, $(1i,2g,3d,3s)$
(middle), and (c) $(1l,2j,3h,4f,5p)$, $(1m,2k,3i,4g,5d,6s)$ (right) levels respectively, of 3D isotropic CHO as function 
of cavity radius.}
\end{figure}

Energy variations of Tables I, II are vividly depicted in Figure 1. The left panel (a) gives such plots for ten 
$s$-waves ($\ell=0$) having radial quantum numbers $n=0-9$, while in right panel (b), we consider same for ten 
lowest states (having $\ell=0-9$ values) corresponding to the lowest $n$ quantum number. Energy axes in both 
cases cover same range; whereas $r_c$ axes differ in two occasions. In both cases, however, confinement in very 
small box radius is ignored to avoid high energy values, which makes these plots difficult to comprehend. To the best of our 
knowledge, such energy plots have been attempted before \cite{montgomery10} only for ground state for a confinement 
region of $r_c=0.5-3.5$ a.u. Present plots significantly extend the radius of sphere, and also cover low, high excited states, 
corroborating the previous findings for ground state. In both panels, neighboring plots are parallel and remain well 
separated. Generally, at low $r_c$, energies assume high values, falling off sharply with an increase in $r_c$ to approach
the isotropic FHO results and finally becomes constant after attaining those. This is in keeping with the fact that energy
of a confining system possessing spherical symmetry is a decreasing monotonic function of confining radius and as $r_c$ is
increased energy levels smoothly converge to the corresponding free system energy levels \cite{stevanovic08a}.

At this stage, we discuss the correlation in energy between 3D isotropic CHO and its unconfined counterpart FHO. It is well
known \cite{cohen92} that the 3D isotropic FHO energy levels, given by $E_{k,\ell}=(k+\ell+\frac{3}{2})=(m+\frac{3}{2})$ a.u.,
are degenerate. Here $k$ is zero or \emph{even} positive integer, $\ell$ is zero or 
\emph{any} positive integer, so that $m$ can have \emph{all} integral values, zero or positive. Thus $\ell,m$ quantum numbers
have same parity. Interestingly, when the 3D isotropic FHO is confined inside a spherical box, not only this characteristic 
degeneracy removed, but also the equal energy separation between adjacent 
levels of 3D FHO disappears \cite{aquino97, stevanovic08}. In left panel (a) of Figure 2, is shown the degeneracy 
breaking of three lowest degenerate levels of isotropic 3D FHO in presence of confinement, \emph{viz.,} $(1d, 2s)$, $(1f, 2p)$, 
$(1g,2d,3s)$ having degeneracies of 2, 2, 3 respectively. For smaller values of $r_c$, the spectrum is clearly non-degenerate;
as the former gradually increased, energy differences become smaller and finally at sufficiently high $r_c$, the levels merge,
making them degenerate. However, in all these three instances, the three degenerate levels remain well separated and do not mix 
with each other. In middle panel (b) is given the degeneracy breaking of two $(1h,2f,3p)$, $(1i,2g,3d,3s)$ levels of FHO
(both having degeneracy of 3) under the influence of confinement. In this case, one notices moderate mixing of $1i$ 
and $3p$ states. Such mixings also happen between $1h$ and $3s$ states (not shown). Finally, in right panel (c) such crossings 
are shown between the first penta-degenerate $(1l,2j,3h,4f,5p)$ and hexa-degenerate $(1m,2k,3i,4g,5d,6s)$ levels respectively of the 
FHO. It has been pointed out that, when energy levels are functions of radius of confinement, it may so happen that, for some 
particular values of $r_c$, energies of two 3D CHO states coincide. This has been termed as \emph{accidental degeneracy} in the 
sense it is defined in \cite{elliott79, stevanovic08a}. The above mixing characteristics are manifestations of such intersections
which happens for $(1h, 3s)$, $(1i, 3p)$, $(1j, 4s)$, $(1j, 3d)$, $(1j, 2g)$ pairs. The second one can be seen from (b), while in 
(c) several such cases appear. Clearly one encounters many complex splitting of energy levels for high $n,\ell$ quantum numbers 
as the box radius is decreased. In all these plots, as in Figure 1, energies in low-$r_c$ region is omitted for ease of 
appreciation. 

\begingroup
\squeezetable
\begin{table}
\caption {\label{tab:table3}Selected expectation values (in a.u.) of some low-lying states of 3D isotropic CHO at 
various $r_c$ values. Numbers in the parentheses denote literature results.} 
\begin{ruledtabular}
\begin{tabular}{lllcccc}
$r_c$  & $n$   &  $\ell$ & $\langle r^{-2} \rangle $ &  $\langle r^{-1} \rangle $  &  $\langle r \rangle $ &  $\langle r^2 \rangle $   \\
\hline
0.5    &   0    &  0   &  35.669782840   &  4.8770025194               &  0.24993143144        &  0.07063332532        \\
       &        &      &  (35.670362626\footnotemark[1])               & (4.8770025195\footnotemark[2])    
                       &  (0.24993143143\footnotemark[2])              &  (0.07063332531\footnotemark[1])       \\
       &   1    &  0   &  75.007610219(75.00992708\footnotemark[1]) &  6.2285137551 &  0.25002044046  
                       & 0.08017671929(0.08017667193\footnotemark[1])        \\
       &   0    &  1   & 15.604579050(15.604579051\footnotemark[1]     & 3.7023250014          & 0.29578935362   
                       & 0.09362593573(0.09362593572\footnotemark[1])        \\ 
       &   1    &  1   & 28.8493991749(28.8493991689\footnotemark[1])  & 4.6995657221          & 0.26969402489   
                       & 0.08682928611(0.08682928612\footnotemark[1])          \\ 
1.0    &   0    &  0   & 9.0239897183(9.0241006941\footnotemark[1])    & 2.4512273473          & 0.49781081198  
                       & 0.28044919195(0.2804491919\footnotemark[1])           \\  
       &   1    &  0   & 18.773367559(18.773803815\footnotemark[1])    & 3.1127583237          & 0.50064924778
                       & 0.32128003811(0.3212800381\footnotemark[1])            \\
       &   0    &  1   & 3.9247112833(3.9247112835\footnotemark[1])    & 1.8561980355          & 0.59025087969 
                       & 0.37296446428(0.3729644642\footnotemark[1])            \\
       &   1    &  1   & 7.2212001400(7.2212001399\footnotemark[1])    & 2.3501690242          & 0.53961515964 
                       & 0.34762136892(0.3476213689\footnotemark[1])            \\
\end{tabular}
\end{ruledtabular}
\begin{tabbing}
$^{\mathrm{a}}$Ref.~\cite{aquino97}. \hspace{60pt}  \=
$^{\mathrm{b}}$Ref.~\cite{montgomery10}. 
\end{tabbing}
\end{table}
\endgroup

Energy orderings in 3D isotropic CHO states in the limit of $r_c \rightarrow \infty$ is found to be, 
\begin{multline}
 1s,1p,(1d,2s),(1f,2p),(1g,2d,3s),(1h,2f,3p),(1i,2g,3d,4s),(1j,2h, 3f,4p), \\  \nonumber
 (1k,2i,3g,4d,5s),(1l,2j,3h,4f,5p),(1m,2k,3i,4g,5d,6s), \cdots \nonumber 
\end{multline}
For convenience, the degenerate states are grouped in parentheses.
Such orderings up to $4s$ were noticed in the variational calculation of \cite{aquino97} and up to $1j$ in \cite{stevanovic08a} 
from the zeros of Bessel functions. Our calculation reproduces these two previous works and present an extended ordering
of levels. In the limit of $r_c \rightarrow 0$, the same changes to, 
\[ 1s,1p,1d,2s,1f,2p,1g,2d,1h,3s,2f,1i,3p,1j,2g,3d,4s,1k,2h,3f,1l,4p,2i, \cdots \]

\begin{figure}
\begin{minipage}[c]{0.28\textwidth}
\centering
\includegraphics[scale=0.32]{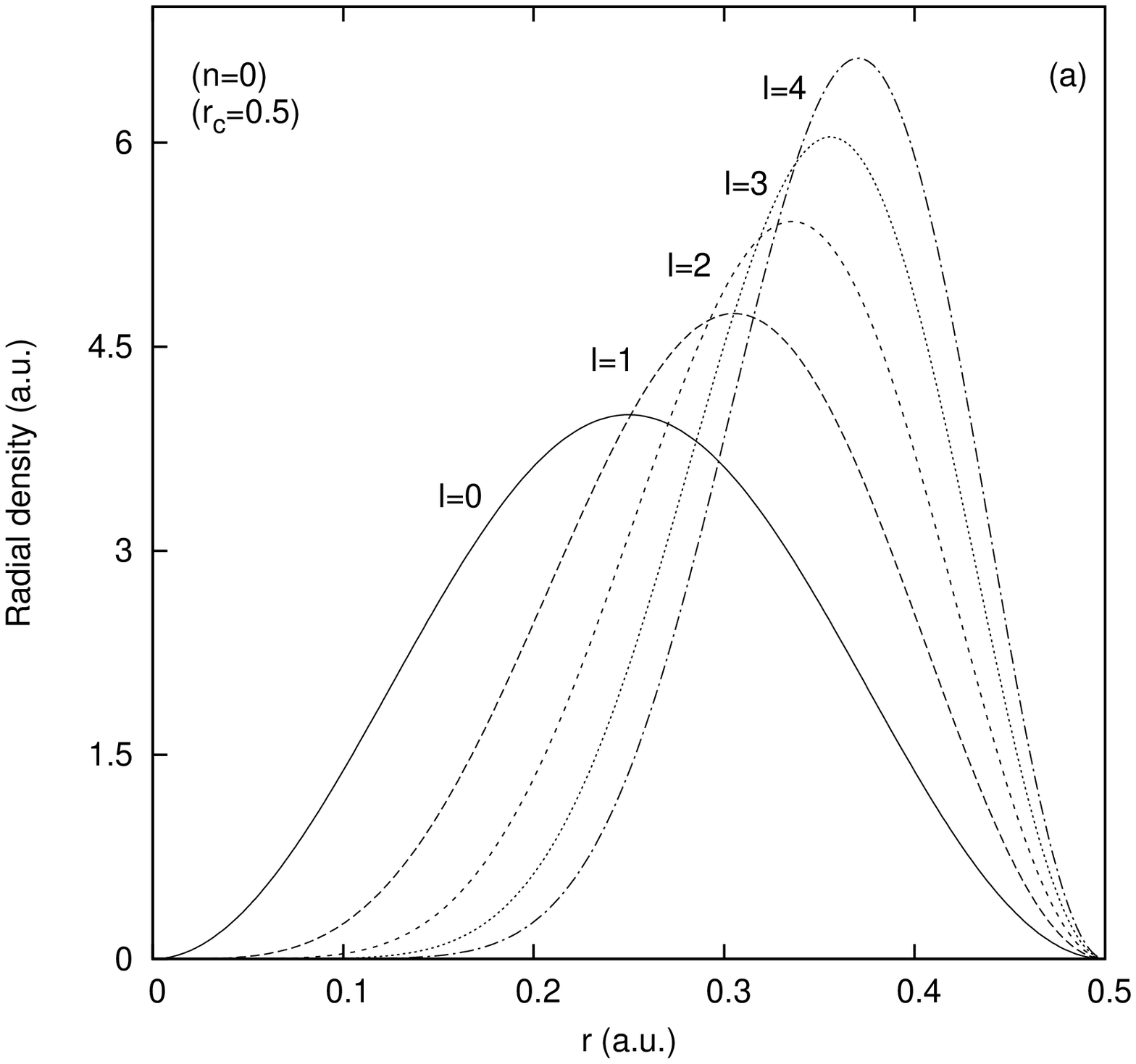}
\end{minipage}
\hspace{0.3in}
\begin{minipage}[c]{0.28\textwidth}
\centering
\includegraphics[scale=0.32]{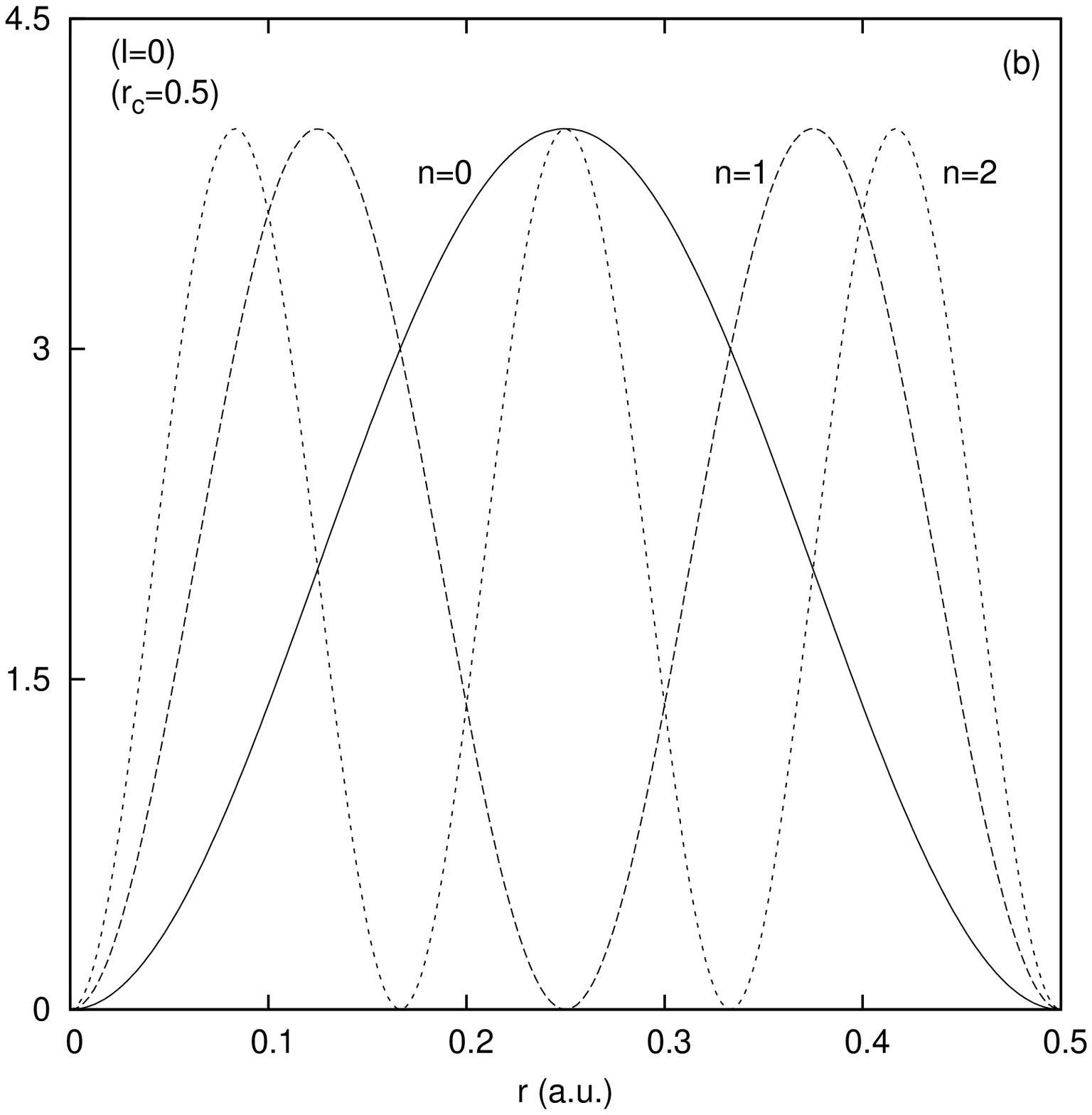}
\end{minipage}
\hspace{0.3in}
\begin{minipage}[c]{0.28\textwidth}
\centering
\includegraphics[scale=0.32]{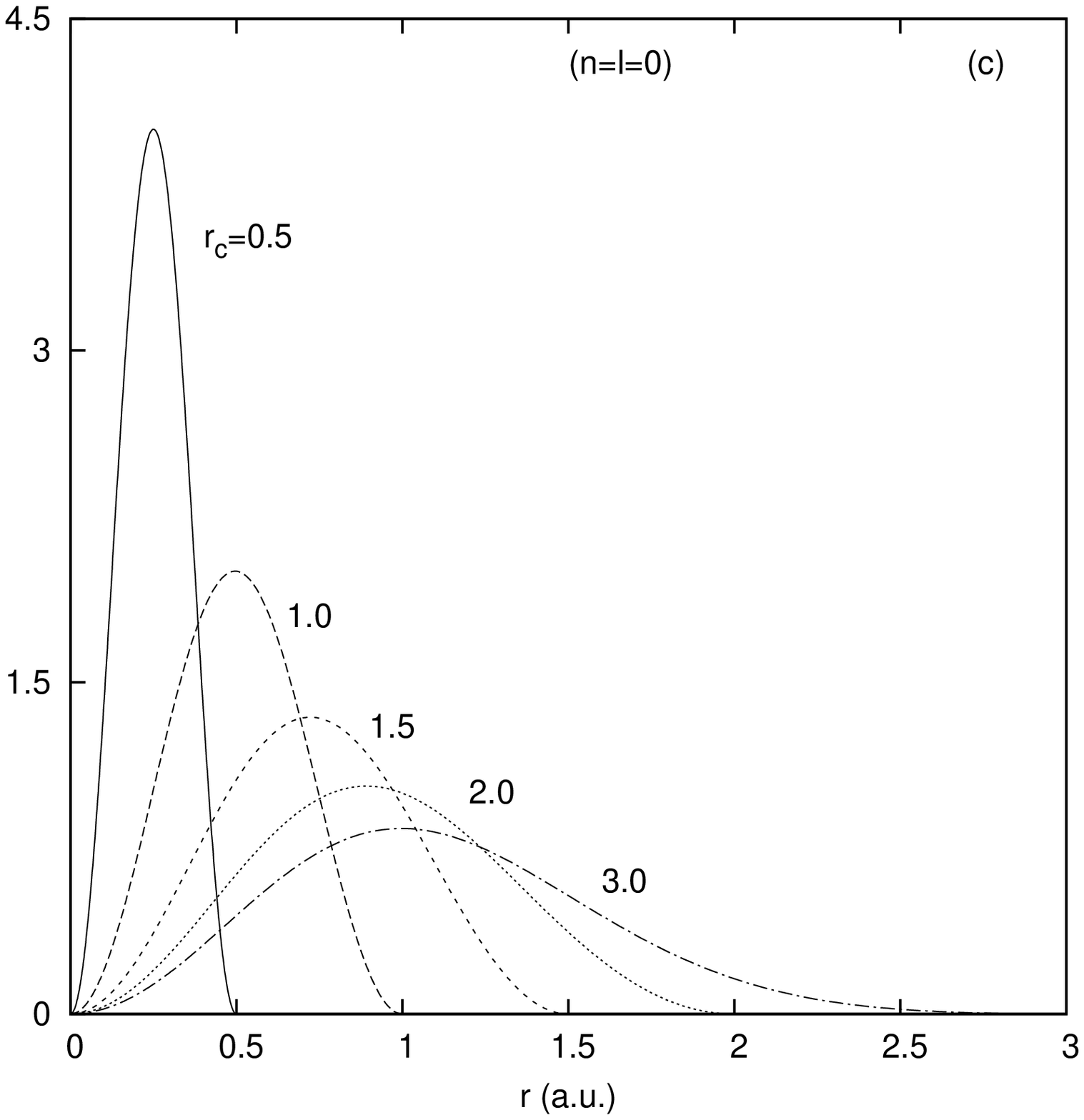}
\end{minipage}
\caption{Radial distribution functions (a.u.) of 3D isotropic CHO: (a) $n=0; l=0-4$ states for $r_c=0.5$; (b) $l=0; n=0-2$ 
states for $r_c=0.5$; (c) ground state for $r_c=0.5,1.0,1.5,2.0,3.0$.}
\end{figure}

\noindent
In this case also, the relative ordering of first 13 states given in \cite{aquino97} and 17 states considered in 
\cite{stevanovic08a} are the same as found here, although we have covered some higher states. 

As a further proof of the usefulness of present work, Table III reports four position expectation values, \emph{viz.}, 
$\langle r^{-2} \rangle$, $\langle r^{-1} \rangle$, $\langle r \rangle$, $\langle r^2 \rangle$ of 3D isotropic CHO
under the spherical confinement. Two lowest $n$ states (0 and 1) corresponding to two lowest $\ell$ values of 0 and 1 are
given for two $r_c$ (0.5, 1). The first and last expectation values for all the states were earlier reported quite accurately 
in \cite{aquino97}. 
Our results show very good agreement with these reference values, which is apparently better for latter than the former. 
The other two expectation values, $\langle r^{-1} \rangle$, $\langle r \rangle$ are available only for ground state having an 
$r_c=0.5$. Once again, GPS values are in excellent agreement with the literature values. No reference results could be found 
for these two expectation values for remaining 7 states.

\begingroup
\squeezetable
\begin{table}
\caption {\label{tab:table4} Lowest six eigenvalues (in a.u.) of 3D confined isotropic quartic oscillator at selected 
$r_c$ values. Ground-state energy of the free oscillator is 1.8998365149 a.u. \cite{taseli97}.} 
\begin{ruledtabular}
\begin{tabular}{l|llllll}
$r_c$  & E$_{0,0}$ & E$_{0,1}$       &  E$_{0,2}$     &  E$_{1,0}$      &   E$_{0,3}$     &  E$_{1,1}$   \\
\hline
0.1   &  493.48022579  &  1009.5364365   &  1660.8731068   &  1973.9208888   &  2441.5596954   &  2983.9758067  \\
0.3   &  54.831597579  &  112.17141900   &  184.54236135   &  219.32525354   &  271.28548566   &  331.55364336  \\
0.5   &  19.742773482  &  40.386895144   &  66.441916477   &  78.962323706   &  97.670693709   &  119.36502732  \\
0.8   &  7.7339639376  &  15.809618718   &  25.996944579   &  30.878482788   &  38.203785113   &  46.663916029  \\
1.0   &  4.9915845102  &  10.182113375   &  16.720373986   &  19.827020261   &  24.548286154   &  29.935709434  \\
1.3   &  3.0794268551  &  6.2185008173   &  10.143870741   &  11.930708685   &  14.823708103   &  17.930956677  \\
1.5   &  2.4678149050  &  4.9124620930   &  7.9342161030   &  9.2167903356   &  11.511774354   &  13.748897606  \\
2.0   &  1.9412531888  &  3.6821177561   &  5.7125440986   &  6.2917841210   &  8.0174364761   &  8.9903956010  \\
2.5   &  1.9002398159  &  3.5559309679   &  5.4265578434   &  5.8330487239   &  7.4754923256   &  8.0492393137  \\
3.0   &  1.8998367875  &  3.5542235225   &  5.4212247115   &  5.8223857288   &  7.4617247879   &  8.0163868627  \\
5.0   &  1.8998365150  &  3.5542220840   &  5.4212190000   &  5.8223727555   &  7.4617057604   &  8.0163311984  \\
\end{tabular}
\end{ruledtabular}
\end{table}
\endgroup

Now, Fig.~3 depicts some sample radial density distribution functions in case of isotropic bounded HO. In left panel (a), 
we show such plots for 5 lowest $n=0$ states having $l$ values 0--4 confined in a box of radius $r_c=0.5$. As $l$ increases, 
the peak height increases, shifts to the right side and tends to become more compact. The middle panel (b) displays 
density plots for 3 lowest states (having $n =0-2$) corresponding to $l=0$ for same fixed $r_c$ as in (a). With an increase 
in $n$, one notices an increase in the number of nodes, whereas the maxima in density peaks remain practically unchanged. 
Finally (c) shows density changes in ground state of 3D CHO for five values of box radii, \emph{viz.}, 0.5, 
1.0, 1.5, 2.0 and 3.0 covering a decent range of confinement. With increase in $r_c$, density peaks reduce to smaller 
values and tend to spread, eventually assuming the shape of that of an FHO for sufficiently large $r_c$. Number of 
nodes of a state remains same for all values of $r_c$. 

Next we consider confinement in case of an isotropic 3D quartic harmonic oscillator. Six lowest eigenstates of this 
system are tabulated in Table IV for various box radii $r_c$. Here too a broad range of $r_c$ is considered.
To the best of our knowledge, such confinement studies have not been presented before. However, some eigenstates in free system 
have been reported in \cite{taseli97}, where extremely accurate energies were obtained by expanding the wave function in a 
Fourier-Bessel series. Our ground-state energy of 1.8998365150 a.u. at last $r_c$ is in very good agreement with 
their estimated value of 1.89988365149 a.u. Note that in this case, the critical boundary parameter $L_{cr}$, according to
\cite{taseli97} is 4.75. This is defined as the value of the boundary length for which $\mathrm{E(L_{cr})- E(L_{cr}+\delta)}$ 
remains below a certain predefined small value (they fixed at $10^{-30}$), for all values of $\delta > 0$. The energy 
orderings (both unconfined and confined) in this case also follow same sequence as found for 3D isotropic harmonic oscillator 
(this is confirmed for all the states given for harmonic oscillator). Of course, in the free quartic oscillator case, the 
degeneracies now disappear.

Now, Fig. 4 displays the above energy variations of 3D quartic isotropic oscillator with respect to changes in box radius.
As in Fig.~1, here also, first ten $l=0$ states corresponding to $n=0-9$ are shown in left panel (a), whereas the right panel (b) 
considers first 
ten $n=0$ states with angular quantum number 0--9. The $r_c$ axis is kept fixed in both cases, while energy axis differs. 
Very small confinements are again omitted for easy appreciation of the plots. Qualitatively speaking the general trend remains
very similar to those in Fig. 1 for 3D CHO. Adjacent curves remain well-separated and parallel in both occasions. Energies tend 
to fall off sharply with an increase in $r_c$ at the beginning; then becomes stationary after reaching the value of that of 
the respective 3D unconfined oscillator. One finds that, free unconfined oscillator energies are attained 
for relatively smaller values of $r_c$ in this case, compared to the same in CHO. Such plots are given here for first time. 

Next we shift our focus on confinement studies in 3D polynomial potentials of higher degree, namely $2K$, with $K=3-10$. At first, 
however, Table V provides six eigenstates of respective free oscillators corresponding to $n,\ell$ quantum numbers 
(0,5) and (0,5,10) respectively, i.e., (0,0), (0,5), (0,10), (5,0), (5,5), (5,10) states. In this case, literature data is 
rather scanty; we could find only a lone reference of \cite{taseli97}. The authors reported first five of above states of
anharmonic oscillators having $K=3,4,10$. In all occasions, GPS results are identical to reference energies except for 
lowest state, where they differ from each other by only $2 \times 10^{-10}$ a.u. This clearly demonstrates the validity 
and efficiency of current approach.

\begin{figure}
\begin{minipage}[c]{0.40\textwidth}
\centering
\includegraphics[scale=0.45]{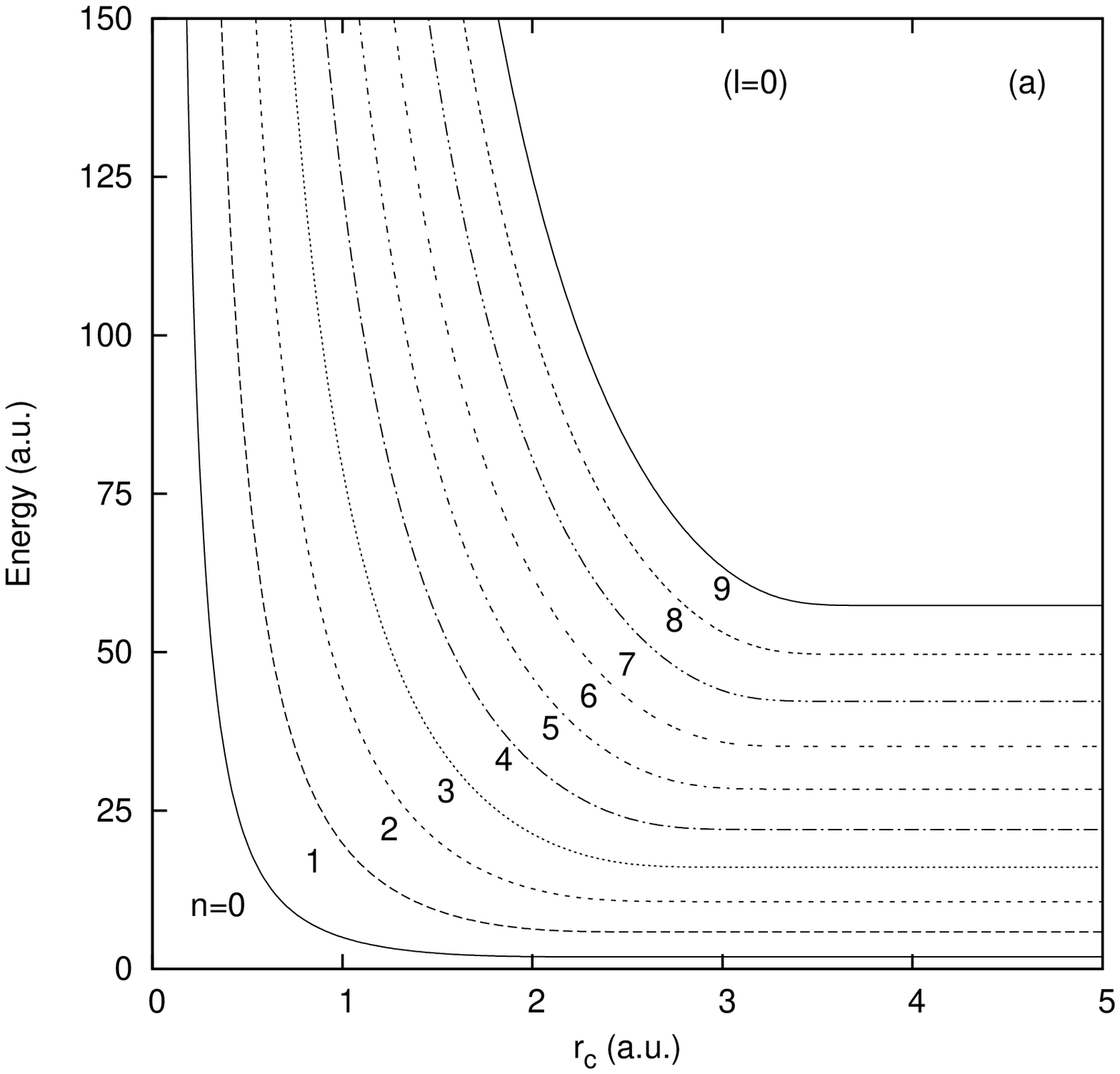}
\end{minipage}%
\hspace{0.5in}
\begin{minipage}[c]{0.40\textwidth}
\centering
\includegraphics[scale=0.45]{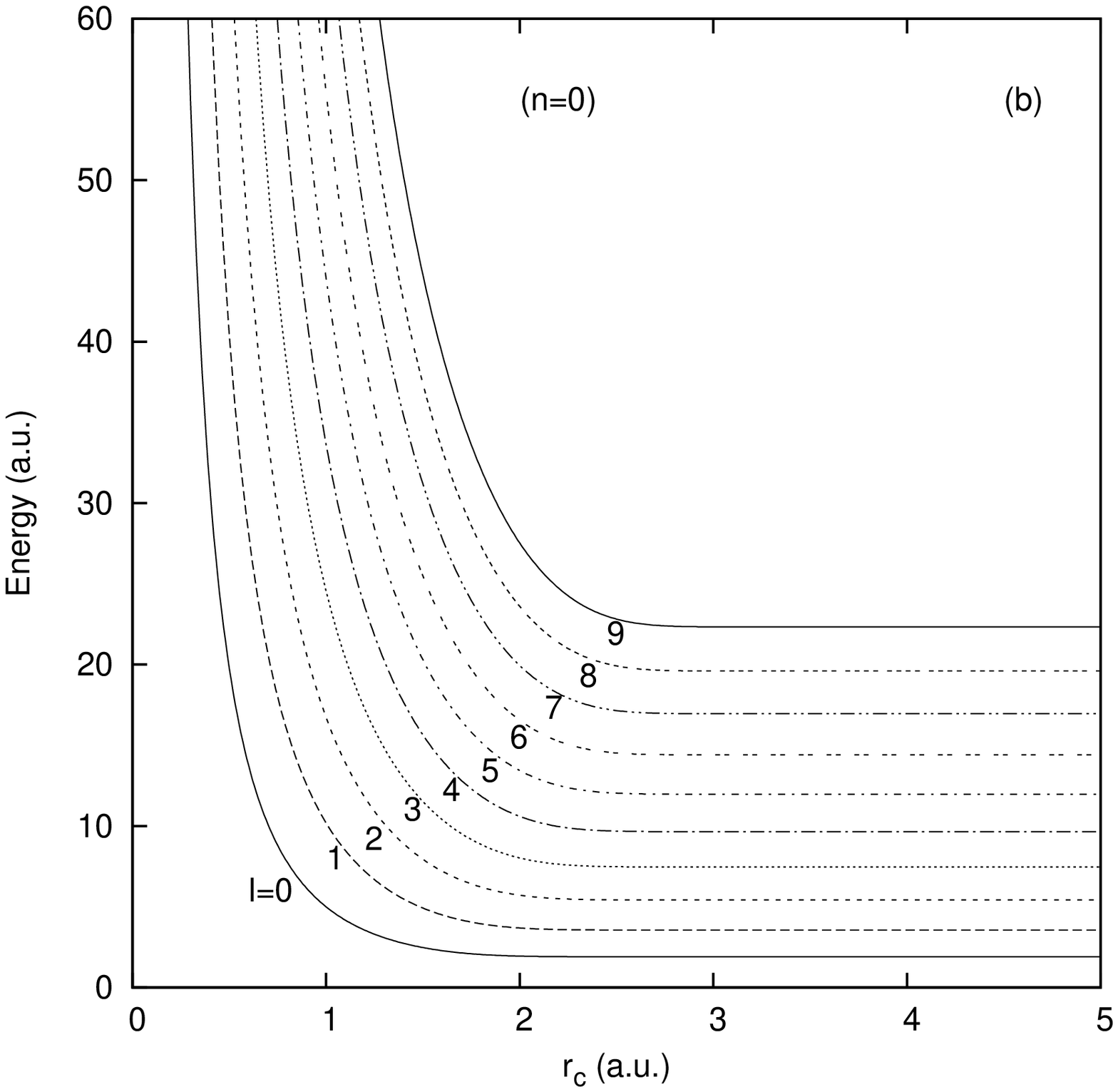}
\end{minipage}%
\caption{Energy changes of 3D confined quartic HO potential for (a) $l=0; n=0-9$ and (b) $n=0; l=0-9$ levels 
respectively, as function of confinement radius $r_c$, in a.u.}
\end{figure}

Finally, we present results of spherical confinement in some high, even-order isotropic 3D anharmonic oscillators. Ground 
states of six such polynomial oscillators having $2K=6,8,10,12,16$ and 20 are offered in Table VI, varying the box radius from 
small to larger values. We are not aware of any such confinement studies before; hence no reference values could be quoted for 
direct comparison. However, the lowest states of free anharmonic oscillators having $2K$ values 6,8 and 20 were estimated  
earlier \cite{taseli97}. Our GPS energies in these three occasions show very good agreement with these reference values. For 
lower $r_c$, energies seem to change rather slowly with increase in $2K$. The estimated values of critical boundaries 
$L_{cr}=3.6, 2.9$ and 1.72 in these three states \cite{taseli97}, decrease with the power and are well supported by energies given
in this table. Considering the performance of our present approach in the context of various confinement situations discussed 
earlier, we are confident that these results of higher order polynomials would also be equally accurate and reliable. Although
we have not ventured into estimating the $L_{cr}$, in general, however, the present calculation seems to corroborate the results 
of \cite{taseli97}. 

\begingroup
\squeezetable
\begin{table}
\caption {\label{tab:table5}Some low- and high-lying eigenvalues (in a.u.) of unconfined 3D polynomial potentials of even 
degree (2K). Literature values in parentheses are quoted from \cite{taseli97}.} 
\begin{ruledtabular}
\begin{tabular}{lllllll}
$K$  &  E$_{0,0}$    &   E$_{0,5}$     &   E$_{0,10}$     &  E$_{5,0}$      &   E$_{5,5}$     &   E$_{5,10}$   \\
\hline
3  &  2.1692993559   &  15.859640450   &  36.010035895    &  44.196187884   &  74.182455900   &  107.23827462  \\
   & (2.1692993557)  & (15.859640450)  & (36.010035895)   & (44.196187884)  & (74.182455900)  &                \\
4  &  2.3779372071   &  18.693357492   &  44.256045884    &  57.560400554   &  99.689155552   &  147.06867947  \\   
   & (2.3779372069)  & (18.693357492)  & (44.256045884)   & (57.560400554)  & (99.689155552)  &                \\   
5  &  2.5489382648   &  20.856314303   &  50.629764312    &  68.606297125   &  121.31727693   &  181.41445326  \\
6  &  2.6934707830   &  22.574874381   &  55.691777381    &  77.756594924   &  139.54557480   &  210.69341741  \\  
7  &  2.8180925155   &  23.984159530   &  59.816772453    &  85.407643316   &  154.97516807   &  235.68085749  \\ 
8  &  2.9271124526   &  25.168625012   &  63.253598497    &  91.876672206   &  168.13894256   &  257.12973408  \\    
9  &  3.0235635760   &  26.183608461   &  66.170973123    &  97.407233060   &  179.46908706   &  275.67781244  \\
10 &  3.1096802645   &  27.066934477   &  68.686384500    &  102.18488145   &  189.30659360   &  291.84162253  \\ 
   & (3.1096802643)  & (27.066934477)  & (68.686384500)   & (102.18488145)  & (189.30659360)  &                \\ 
\end{tabular}
\end{ruledtabular}
\end{table}
\endgroup

These above features of previous table are well reflected in Fig. 5, where energies of three low-lying states, namely, (0,0), 
(0,1) and (1,0), of 3D spherically confined isotropic anharmonic oscillators of even orders $2,4,6, \cdots, 20$ are plotted 
against the radius of confinement. Energy ranges are different in three plots while $r_c$ axis remains same in (b), (c). From 
very close values at smaller $r_c$, individual energies decrease sharply and then assumes the constant value of corresponding
unconfined potential, for critical values of the
boundary parameter $L_{cr}$. As evident, in all these three occasions, $L_{cr}$ decrease as $2K$ increase.

\begingroup
\squeezetable
\begin{table}
\caption {\label{tab:table6} Ground-state energies (in a.u.) of some higher degree ($2K=6,8,10,12,16,20$) 3D polynomial oscillators 
enclosed within a spherical box of radius $r_c$ a.u.} 
\begin{ruledtabular}
\begin{tabular}{lllllll}
$r_c$  &     $K=3\footnotemark[1]$     &        $K=4\footnotemark[1]$    &       $K=5$     &      $K=6$      
       &       $K=8$     &        $K=10\footnotemark[1]$    \\
\hline
0.1   &  493.48022012  &  493.48022009   &  493.48022009   &  493.48022009   &  493.48022009   &  493.48022009    \\
0.3   &  54.831156039  &  54.831136597   &  54.831135622   &  54.831135569   &  54.831135565   &  54.831135565    \\ 
0.5   &  19.739647612  &  19.739270202   &  19.739218199   &  19.739210339   &  19.739208850   &  19.739208805    \\
0.8   &  7.7179857725  &  7.7132646295   &  7.7116613437   &  7.7110607571   &  7.7107156751   &  7.7106485219    \\      
1.0   &  4.9627782237  &  4.9504713676   &  4.9443993675   &  4.9410809804   &  4.9378971947   &  4.9365423998    \\ 
1.5   &  2.4843022335  &  2.5349157915   &  2.6136557597   &  2.7133768531   &  2.9275486225   &  3.1096805499    \\
2.0   &  2.1713945253  &  2.3779504727   &  2.5489382668   &  2.6934707830   &  2.9271124526   &  3.1096802645    \\
2.5   &  2.1692993821  &  2.3779372071   &  2.5489382648   &  2.6934707830   &  2.9271124526   &  3.1096802645    \\
3.0   &  2.1692993559  &  2.3779372071   &  2.5489382648   &  2.6934707830   &  2.9271124526   &  3.1096802645    \\
4.0   &  2.1692993559  &  2.3779372071   &  2.5489382648   &  2.6934707830   &  2.9271124526   &  3.1096802645    \\
\end{tabular}
\end{ruledtabular}
\begin{tabbing}
$^{\mathrm{a}}$ Energies at $r_c \rightarrow \infty$ for $K=3,4,10$ respectively are: 2.1692993558, 2.3779372070 and 
3.1096802643 a.u. \cite{taseli97}.
\end{tabbing}
\end{table}
\endgroup

\section{conclusion}
A detailed analysis has been made to understand confinement in the family of 3D anharmonic oscillators of even order enclosed
within spherically impenetrable walls. Energies, expectation values, radial densities are obtained by means of GPS method. 
For isotropic 3D CHO, these quantities show excellent agreement with the best reference results available in literature, 
covering \emph{small, intermediate and large} radius of confinement. It may be noted that, good-quality results in confined 
systems are obtainable by a number of attractive and elegant methodologies for \emph{medium and large} $r_c$; however, same 
for small $r_c$ is rather scarce. In this work, we are able to generate quite good results in both these regions with equal 
ease and efficacy. Also, similar accuracy results are also offered in case of high-lying states. A thorough analysis of energy 
changes with respect to confinement radius has been made. Energy ordering in case of 3D CHO and free oscillator, as well as 
the degeneracy pattern is also discussed. Several interesting mixings have been pointed out. Next, we perform similar 
investigation on 3D quartic oscillator. Finally we briefly touch upon the case of 3D isotropic even-order isotropic bounded
oscillators for higher orders up to as high as 20. Once again, low- and high-lying states in the confined system and free 
oscillator are presented in some detail. Energy changes with variations in $r_c$ are monitored for selected low-lying states 
in such oscillators. The method also delivers excellent quality results in case of unconfined oscillators. Comparison with 
literature data reveals very good agreement in case of small as well as large box sizes. Confinement studies in other physical 
and chemical situations such as the H atom or many-electron atoms may further consolidate the success of this approach, some of 
which may be taken up in future works. Many new states are reported here for the first time. In summary, a simple, reliable 
method is offered for accurate calculation of harmonic and other higher order polynomial oscillators under spherical confinement.

\begin{figure}
\begin{minipage}[c]{0.28\textwidth}
\centering
\includegraphics[scale=0.32]{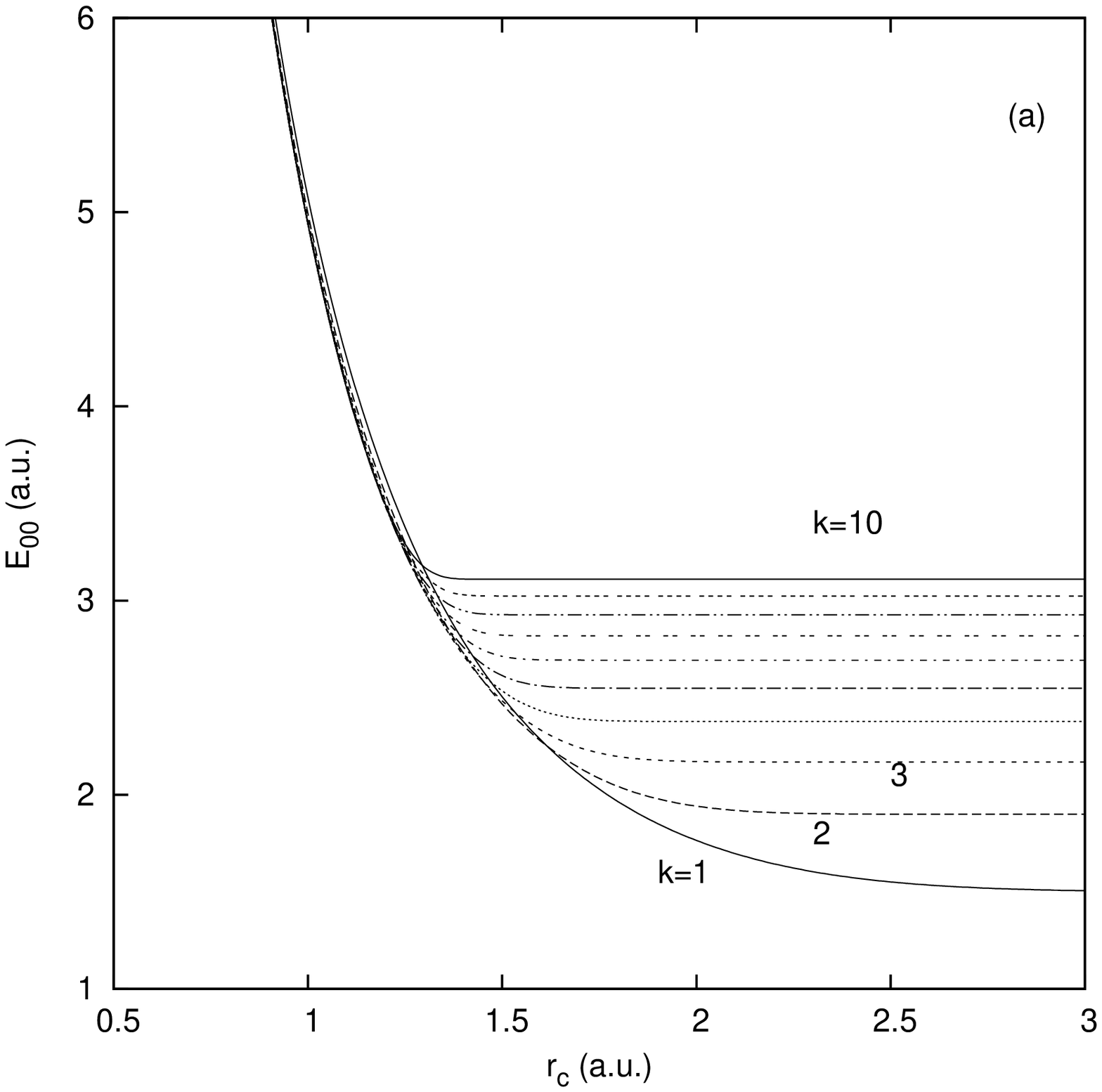}
\end{minipage}
\hspace{0.3in}
\begin{minipage}[c]{0.28\textwidth}
\centering
\includegraphics[scale=0.32]{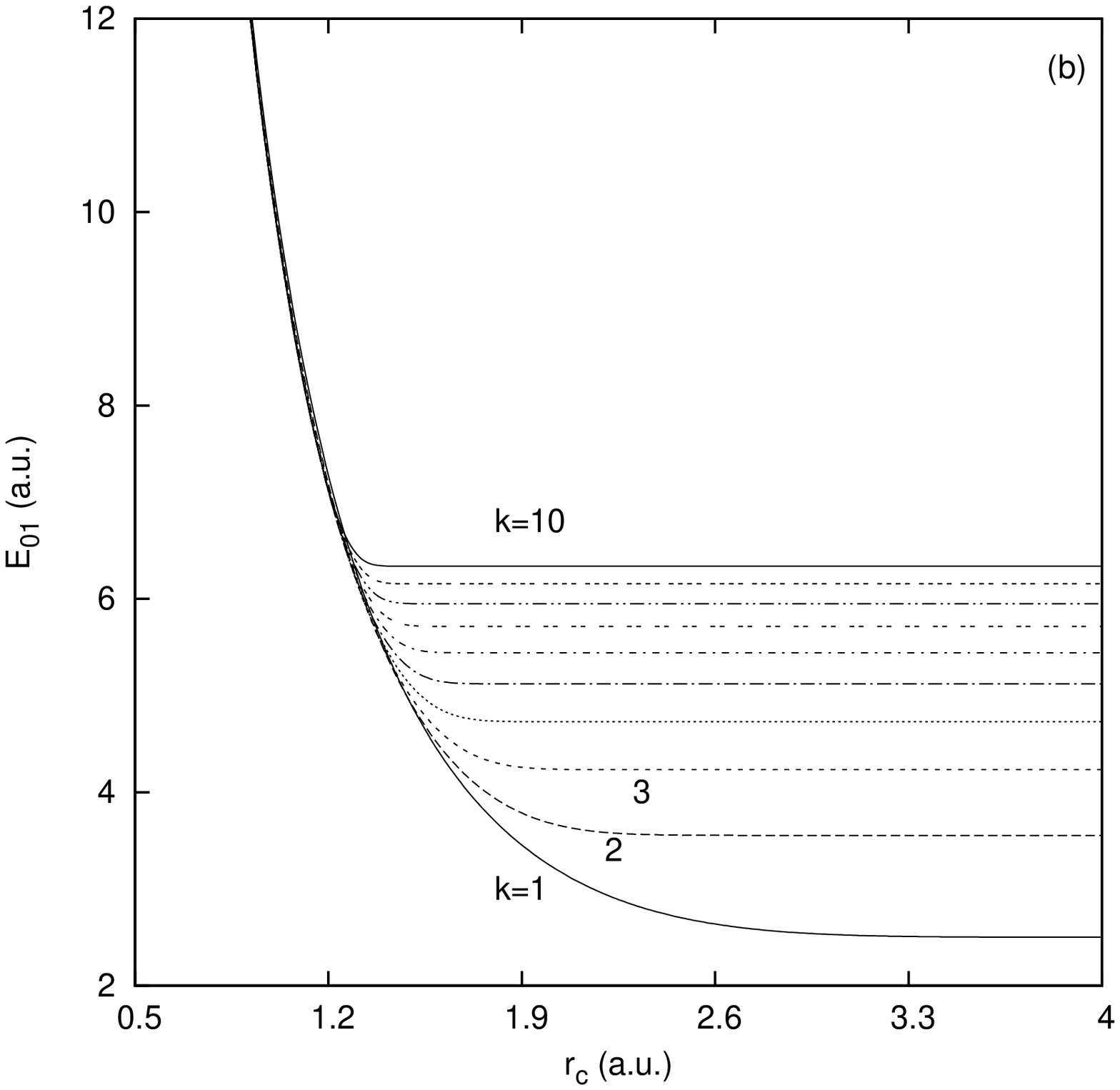}
\end{minipage}
\hspace{0.3in}
\begin{minipage}[c]{0.28\textwidth}
\centering
\includegraphics[scale=0.32]{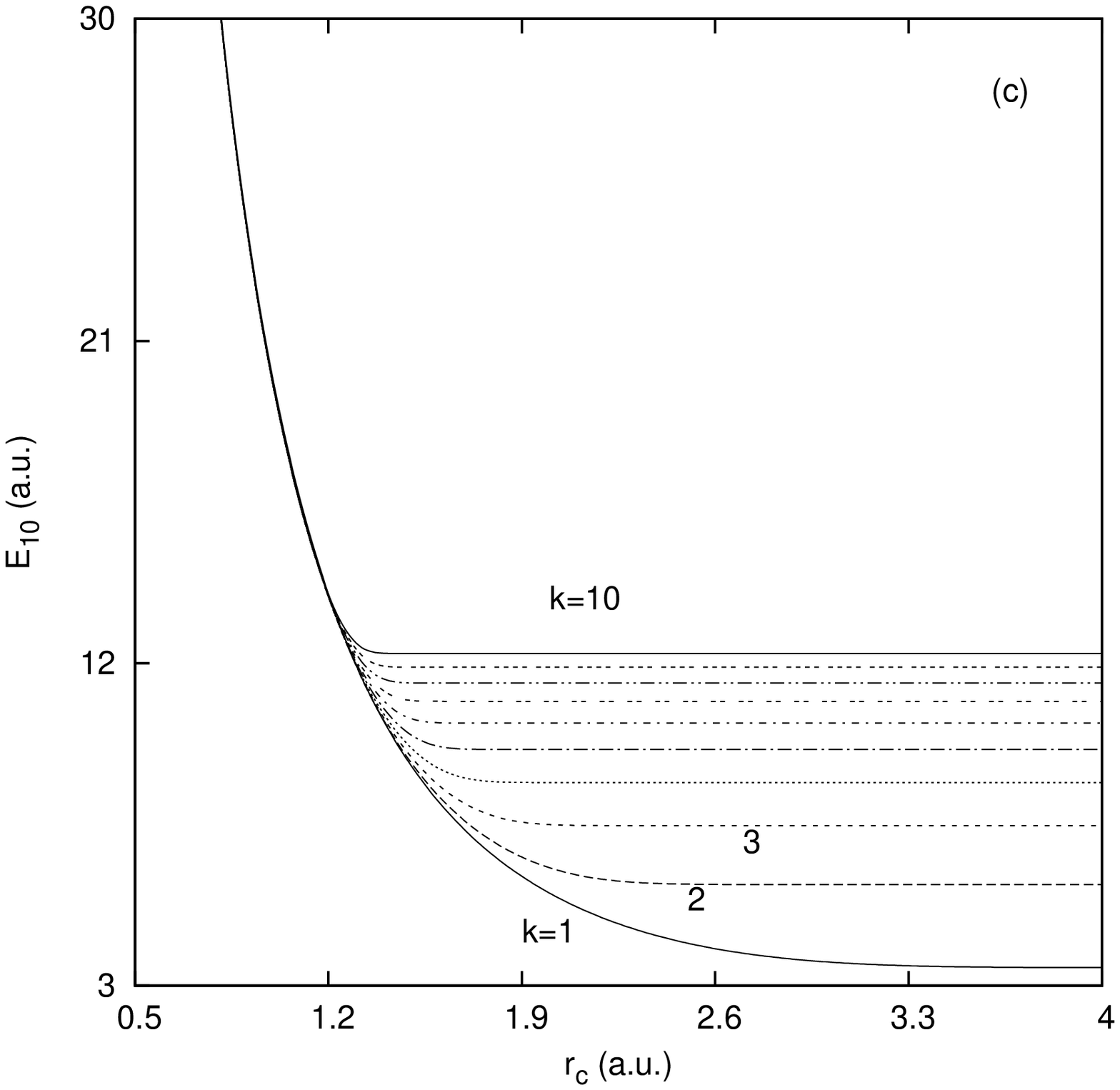}
\end{minipage}
\caption{Energies in spherically confined isotropic 3D even-order anharmonic oscillators ($2K=2,3,4,\cdots,10$). Left, middle and
right panels (a), (b), (c) correspond to E$_{0,0}$, E$_{0,1}$ and E$_{1,0}$ states with respect to box radius $r_c$, respectively.
All quantities are in a.u. See text for details.} 
\end{figure}



\begin{thebibliography}{99}
\bibitem{michels37} A.~Michels, J.~De Boer and A.~Bijli, Physica \textbf{4}, 981 (1937). 
\bibitem{fowler84} P.~W.~Fowler, Mol.~Phys.~ \textbf{53}, 865 (1984); \textit{ibid.} \textbf{54}, 129 (1985). 
\bibitem{froman87} P.~O.~Fr\"oman, S.~Yngve and N.~J.~Fr\"oman, J.~Math.~Phys.~ \textbf{28}, 1813 (1987).
\bibitem{yngve88} S.~Yngve, J.~Math.~Phys.~ \textbf{29}, 931 (1988). 
\bibitem{jaskolski96} W.~Jask\'olski, Phys.~Rep.~ \textbf{271}, 1 (1996). 
\bibitem{connerade00} J.~P.~Connerade, V.~H.~Dolmatov and P.~A.~Lakshmi, J.~Phys.~B \textbf{33}, 251 (2000). 
\bibitem{buchachenko01} A.~L.~Buchachenko, J.~Phys.~Chem.~A \textbf{105}, 5839 (2001). 
\bibitem{gravesen05} J.~Gravesen, M.~Willatzen and L.~L.~Y.~Voon, Phys.~Scr.~ \textbf{72}, 105 (2005).
\bibitem{heiss05} W.~D.~Heiss (Ed.) \textit{Quantum Dots: A Doorway to Nanoscale Physics}, Springer, Berlin (2005). 
\bibitem{vawter68} R.~Vawter, Phys.~Rev.~ \textbf{174}, 749 (1968). 
\bibitem{navarro80} V.~C.~Aguilera-Navarro, E.~Ley Koo and A.~H.~Zimerman, J.~Phys.~A \textbf{13}, 3585 (1980). 
\bibitem{fernandez81} F.~M.~Fern\'andez and E.~A.~Castro, Int.~J.~Quant.~Chem.~ \textbf{20}, 623 (1981).
\bibitem{arteca83} G.~A.~Arteca, S.~A.~Maluendes, F.~M.~Fern\'andez and E.~A.~Castro, Int.~J.~Quant.~Chem.~ \textbf{24}, 
169 (1983).
\bibitem{taseli93} H.~Ta\c{s}eli, Int.~J.~Quant.~Chem.~ \textbf{46}, 319 (1993). 
\bibitem{vargas96} R.~Vargas, J.~Garza and A.~Vela, Phys.~Rev.~E \textbf{53}, 1954 (1996). 
\bibitem{sinha99} A.~Sinha and R.~Roychoudhury, Int.~J.~Quant.~Chem.~ \textbf{73}, 497 (1999). 
\bibitem{aquino01} N.~Aquino, E.~Casta\~{n}o, G.~Campoy and V.~Granados, Eur.~J.~Phys.~ \textbf{22}, 645 (2001).
\bibitem{campoy02} G.~Campoy, N.~Aquino and V.~D.~Granados, J.~Phys.~A \textbf{35}, 4903 (2002).
\bibitem{fernandez81a} F.~M.~Fern\'andez and E.~A.~Castro, Phys.~Rev.~A \textbf{24}, 2883 (1981).
\bibitem{navarro83} V.~C.~Aguilera-Navarro, J.~F.~Gomes, A.~H.~Zimerman and E.~Ley-Koo, J.~Phys.~A \textbf{16}, 2943 (1983).
\bibitem{marin91} J.~L.~Marin and S.~A.~Cruz, Am.~J.~Phys.~ \textbf{59}, 931 (1991). 
\bibitem{aquino97} N.~Aquino, J.~Phys.~A \textbf{30}, 2403 (1997).
\bibitem{taseli97a} H.~Ta\c{s}eli and A.~Zafer, Int.~J.~Quant.~Chem.~ \textbf{61}, 759 (1997). 
\bibitem{taseli97} H.~Ta\c{s}eli and A.~Zafer, Int.~J.~Quant.~Chem.~ \textbf{63}, 935 (1997). 
\bibitem{sinha03} A.~Sinha, J.~Math.~Chem.~ \textbf{34}, 201 (2003). 
\bibitem{filho03} E.~D.~Filho and R.~M.~Ricotta, Phys.~Lett.~A \textbf{320}, 95 (2003). 
\bibitem{sen06} K.~D.Sen and A.~K.~Roy, Phys.~Lett.~A \textbf{357}, 112 (2006).
\bibitem{montgomery07} H.~E.~Montgomery Jr.~, N.~Aquino and K.~D.~Sen, Int.~J.~Quant.~Chem.~ \textbf{107}, 798 (2007).
\bibitem{jaber08} S.~M.~Al-Jaber, Int.~J.~Theor.~Phys.~ \textbf{47}, 1853 (2008). 
\bibitem{stevanovic08a} Lj.~Stevanovi\'{c} and K.~D.~Sen, J.~Phys.~A \textbf{41}, 225002 (2008). 
\bibitem{stevanovic08} Lj.~Stevanovi\'{c} and K.~D.~Sen, J.~Phys.~A \textbf{41}, 265203 (2008). 
\bibitem{montgomery10} H.~E.~Montgomery Jr.~, G.~Campoy and N.~Aquino, Phys.~Scr.~ \textbf{81}, 045010 (2010).
\bibitem{serrano13} F.~A.~Serrano and S.-H.~Dong, Int.~J.~Quant.~Chem.~ \textbf{113} 2282 (2013).
\bibitem{aquino95} N.~Aquino, Int.~J.~Quant.~Chem.~ \textbf{54}, 107 (1995).
\bibitem{burrows06} B.~L.~Burrows and M.~Cohen, Int.~J.~Quant.~Chem.~ \textbf{106}, 478 (2006).
\bibitem{aquino07} N.~Aquino, G.~Campoy and H.~E.~Montgomery Jr.~ Int.~J.~Quant.~Chem.~ \textbf{107}, 1548 (2007).
\bibitem{ciftci09} H.~Ciftci, R.~L.~Hall and N.~Saad, Int.~J.~Quant.~Chem.~ \textbf{109}, 931 (2009).
\bibitem{roy04} A.~K.~Roy, J.~Phys.~G \textbf{30}, 269 (2004). 
\bibitem{roy04a} A.~K.~Roy, J.~Phys.~B \textbf{37}, 4369 (2004); {\it ibid.} \textbf{38}, 1591 (2005).
\bibitem{roy04b} A.~K.~Roy, Phys.~Lett.~A \textbf{321}, 231 (2004).
\bibitem{roy05} A.~K.~Roy, Int.~J.~Quant.~Chem.~\textbf{104}, 861 (2005), {\it ibid.} \textbf{113}, 1503 (2013), 
{\it ibid.} \textbf{114} 383 (2014). 
\bibitem{roy05a} A.~K.~Roy, Pramana--J.~Phys.~ \textbf{65}, 01 (2005). 
\bibitem{roy08} A.~K.~Roy, A.~F.~Jalbout and E.~I.~Proynov, Int.~J.~Quant.~Chem.~ \textbf{108}, 827 (2008). 
\bibitem{cohen92} C.~Cohen-Tannoudji, B.~Diu and F.~Lalo\"{e}, \emph{Quantum Mechanics}, Wiley-VCH, (1992).
\bibitem{elliott79} J.~P.~Elliott and P.~G.~Dawber, \emph{Symmetry in Physics}, London, Macmillan, (1979).
\end{thebibliography}
\end{document}